\documentclass[letterpaper, 11 pt]{article}  
 
\usepackage[left=2.5cm, right=2.5cm, top=3cm, bottom=3cm]{geometry}                     

\usepackage{lipsum}

\newcommand\blfootnote[1]{%
  \begingroup
  \renewcommand\thefootnote{}\footnote{#1}%
  \addtocounter{footnote}{-1}%
  \endgroup
}

\usepackage{subcaption}
\usepackage{mwe}
\usepackage{algorithm}
\usepackage{balance} 
\usepackage{algpseudocode}
\usepackage{xcolor,calc}

\usepackage{graphics}
\usepackage{hyperref}
\usepackage{epstopdf}
\usepackage{graphicx}
\usepackage{amsmath} 
\usepackage{amssymb}  
\usepackage{amsthm}
\usepackage[noadjust]{cite}
\usepackage{color}
\usepackage{enumerate}
\usepackage{multirow}
\usepackage{mathtools}
\usepackage{booktabs}
\usepackage{epstopdf}
\usepackage[normalem]{ulem}

\usepackage{url}
\usepackage{bm}
\usepackage{caption}

\usepackage{booktabs}

\newtheorem{remark}{Remark}

\makeatletter
\newsavebox\myboxA
\newsavebox\myboxB
\newlength\mylenA

\newcommand*\xoverline[2][0.75]{%
    \sbox{\myboxA}{$\m@th#2$}%
    \setbox\myboxB\null
    \ht\myboxB=\ht\myboxA%
    \dp\myboxB=\dp\myboxA%
    \wd\myboxB=#1\wd\myboxA
    \sbox\myboxB{$\m@th\overline{\copy\myboxB}$}
    \setlength\mylenA{\the\wd\myboxA}
    \addtolength\mylenA{-\the\wd\myboxB}%
    \ifdim\wd\myboxB<\wd\myboxA%
       \rlap{\hskip 0.5\mylenA\usebox\myboxB}{\usebox\myboxA}%
    \else
        \hskip -0.5\mylenA\rlap{\usebox\myboxA}{\hskip 0.5\mylenA\usebox\myboxB}%
    \fi}
\makeatother

\usepackage{tabularx}
\usepackage{booktabs}

\newtheorem{definition}{Definition}
\newtheorem{assumption}{Assumption}

\makeatletter
\let\NAT@parse\undefined
\makeatother

\begin{document}

\title{Learning How to Solve ``Bubble Ball"}

\author{Hotae Lee, Monimoy Bujarbaruah, and Francesco Borrelli\blfootnote{Authors are with the MPC Lab, UC Berkeley, Berkeley, CA 94720, USA; Emails:\{hotae.lee, monimoyb, fborrelli\}@berkeley.edu.}
}

\maketitle

\begin{abstract}%
\noindent ``Bubble Ball" is a game built on a 2D physics engine, where a finite set of objects can modify the motion of a bubble-like ball. The objective is to choose the set and the initial configuration of the objects, in order to get the  ball to reach a target flag. The presence of obstacles, friction, contact forces and combinatorial object choices make the game hard to solve. \\

\noindent In this paper, we propose a hierarchical predictive framework which solves Bubble Ball. Geometric, kinematic and dynamic models are used at different levels of the hierarchy. At each level of the game, 
data collected during failed iterations are used to update models at all hierarchical level and converge to a feasible solution to the game.\\

\noindent The proposed approach successfully solves a large set of  Bubble Ball levels within reasonable number of trials (\href{https://github.com/hotae319/BubbleBall}{GitHub}, \href{https://sites.google.com/berkeley.edu/bubble-ball/home}{Videos}).
This proposed framework can also be used to solve other physics-based games, especially with limited training data from human demonstrations. 

\end{abstract}

\section{Introduction}
The field of physics-based games has garnered significant attention for the testing and evaluation of control algorithms/reinforcement learning methods \cite{mnih2015human, heess2015learning, kurach2019google}.
Most approaches to solve such games are model-free, such as Q-learning or policy gradient methods \cite{mnih2015human, lillicrap2015continuous}. 
However, most of these approaches require sizeable training data and are difficult to generalize across multiple game levels with widely varying configurations. The primary reason is the lack of use of intuitive models/motion primitives arising from the notions of physics, such as the effect of contact forces.

On the other hand, using physics driven high fidelity modeling for solving these problems can also be challenging. For example, the contact forces between any two bodies can result in a discontinuous motion evolution. Moreover, the contact forces themselves are the solutions to hard optimization problems, such as a Linear Complementarity problems \cite{todorov2012mujoco,kuindersma2016optimization, hwangbo2018per}. Using such system modeling approach for solving a physic-based puzzle in real-time can become computationally intractable. 

In this paper, we propose a model-based hierarchical approach to overcome the aforementioned challenges. The presence of physics-based models circumvents the shortcomings of model-free methods, while hierarchies allow for the use of low-complexity models for the right prediction objective. We specifically focus on solving the very popular (millions of downloads) game Bubble Ball \cite{BubbleBall2}. The objective of the game is to decide a configuration of a finite set of objects, in order to get a bubble-like ball to reach a target flag following the laws of physics. Each hierarchy of our approach uses simplified abstractions of the game's physics engine for attempting to solve a game level. Data of failed trials  is used at each hierarchical level to update the models and the goal set, and attempt the next trial. 
This is motivated by recent works such as \cite{rosolia2017learning, hewing2020learning, vallon2020data} .
Our contribution is a disciplined solution approach based on three principles:
\begin{itemize}
      \item \textit{Exploiting Locality}: At each trial we restrict our focus to only a specific spatial region of the state-space, where a failed solution from a previous trial needs improvement. By doing so, we reduce the computational complexity of solution search. 
   
   \item \textit{Iterative Learning}: If a trial of the game fails, using the collected game data we improve our approximated solution by updating the local region and the model parameters of the movable objects. 
   
   \item \textit{Exploiting Model Hierarchy}: We use three hierarchies of models. A geometry-based high-level path of the ball connecting the start and target positions provides us a warm start. Then, using kinematic and dynamic models of the movable objects, we solve a tractable optimization problem to place the objects and track the ball’s planned path in any local region.
\end{itemize}
We demonstrate the performance of our algorithm in Section~\ref{sec:result},
by using the  physics simulator of Bubble Ball (the one used by the actual commercial game).
The proposed approach successfully solves a large set of  Bubble Ball levels within reasonable number of trials (\href{https://github.com/hotae319/BubbleBall}{GitHub}, \href{https://sites.google.com/berkeley.edu/bubble-ball/home}{Videos}). 

Our approach is focused on Bubble Ball and the three principles discussed above are (in one form or another and not necessarily all together) already used by different communities to solve complex problems. Nonetheless, we believe that the exercise presented in this paper can be beneficial to the ``learning to control" community as it tries to bring some discipline on how to glue the different hierarchical levels together.
In particular, the same framework that utilizes models of different complexity for each hierarchy and uses a terminal set to a lower level Model Predictive Control  has been already successfully used  in manipulators'  and race-cars' application \cite{vallon2020data, shen2020collision}. 


\section{Introduction to Bubble Ball}\label{sec:formulation}
\begin{figure}[h!]
    \centering
  \includegraphics[width=0.98\columnwidth]{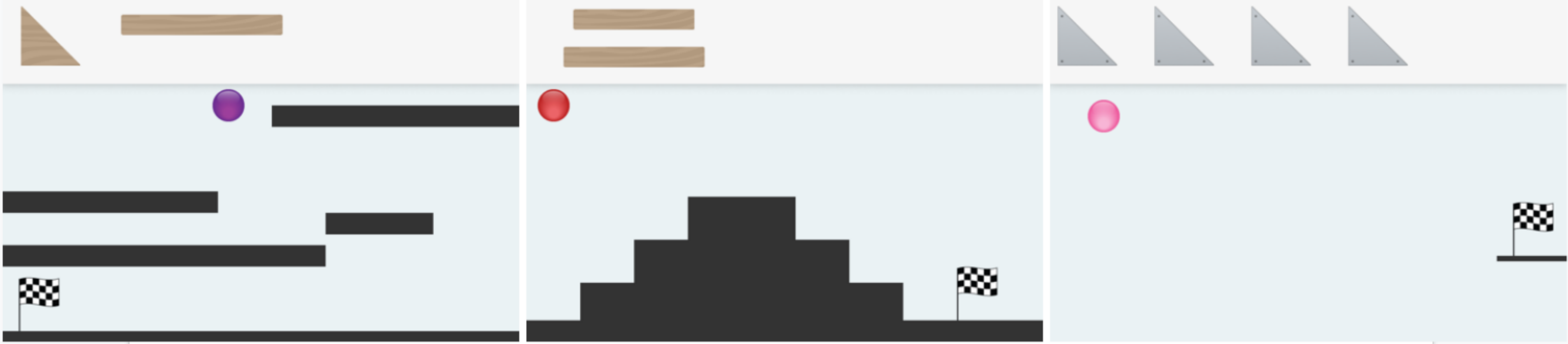}
  \caption{The target position of the ball is marked by the checkered flag. The black objects indicate the environment and the remaining (i.e., wooden and metallic) pieces are the movable objects that the user will place to decide the initial game configuration.}
  \label{fig:bubble_levels}
\end{figure}
Bubble Ball is a game built on a physics engine (Box2D), where the objective is to decide a configuration of a finite set of objects, to get a bubble-like ball to reach a target flag. The game is organized into levels of increasing difficulty. In each level, the player is provided with a different set of movable objects which can affect the ball's trajectory.
Each level also has a set of fixed objects which describe the environment. Fig.~\ref{fig:bubble_levels} shows a set of levels of Bubble Ball depicting the aforementioned components. 

The player decides the initial configuration of a subset of the given movable objects and has no further inputs after the game starts. The trajectories of the ball and the movable objects then evolve under the laws of physics, affected by gravity and contact forces. 
In order to solve the game, the player has to  predict how the objects affect the ball's trajectory. Such prediction is challenging since  contact forces between all the elements of the game during the simulation change discontinuously as a function of the objects' material and the objects' initial state. In addition to the continuous decision variables (the objects' initial configurations), the combinatorial decision on which objects to select adds complexity to the game. 

Human players use multiple trials of the same level to converge to a solution and then advance to the next level. Observations of each unsuccessful trial are then used to modify the previous blocks' initial condition, or to try a different combination of blocks.

\section{Problem Formulation}\label{sec:formulation1}
We model the game evolution as a discrete time nonlinear system interacting with the environment: 
\begin{align}\label{eq:basic_sys}
   &s_{k+1} = f(\begin{bmatrix} s_k \\ m_k \end{bmatrix}, z) ~,~   m_{k+1}=g(\begin{bmatrix} s_k \\ m_k \end{bmatrix}, z),
\end{align}
where $s_k \in \mathbb{R}^4$ is the state of the ball, $m_k \in \mathbb{R}^{8\times n}$ ($n$ objects) is the state of the movable objects, and $z \in \mathbb{R}^{8\times p}$ ($p$ fixed environment objects) is the state of the immovable objects. 
\begin{figure}[th!]
    \centering
  \includegraphics[width=0.45\columnwidth]{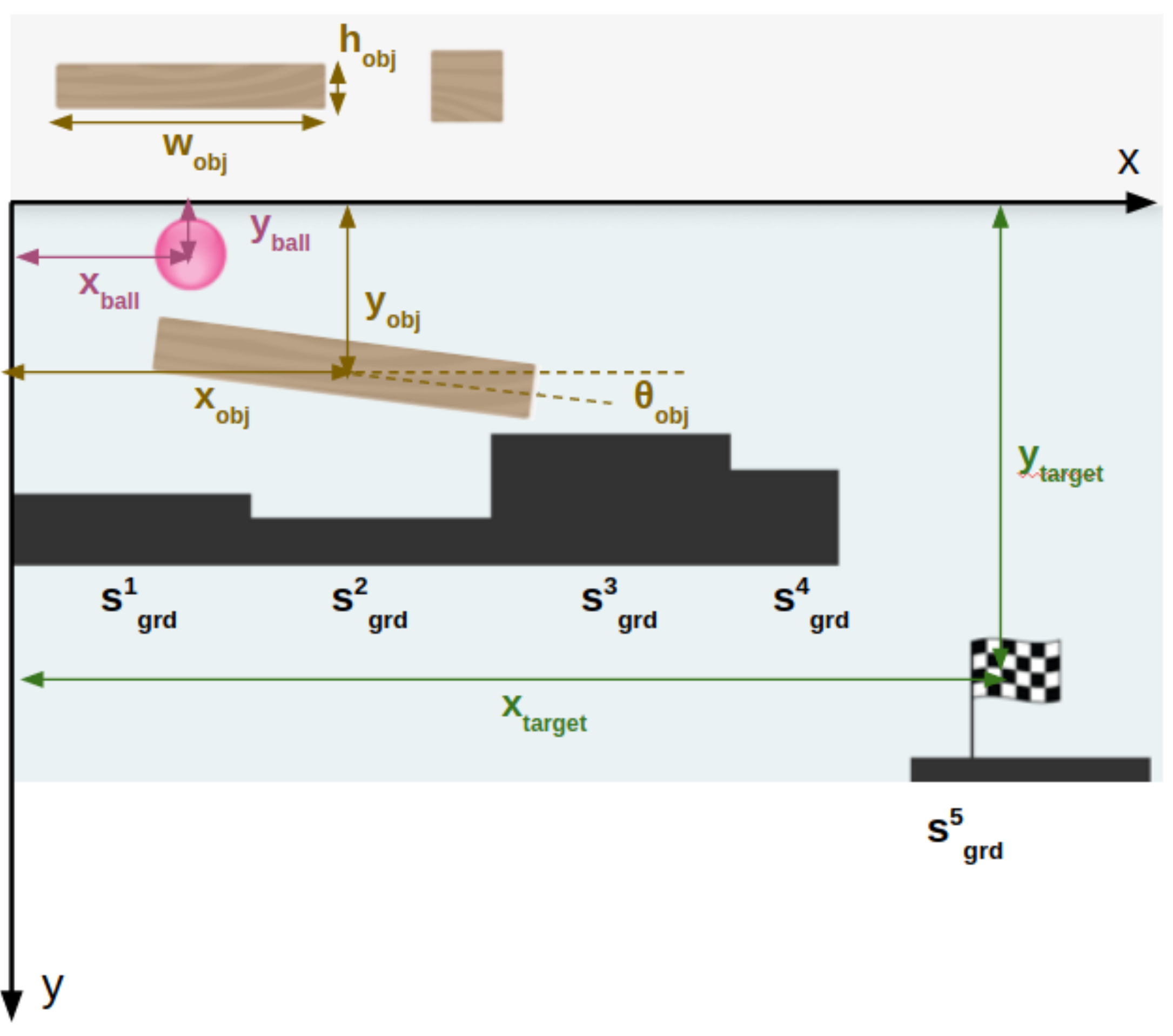}
  \caption{Description of the states and coordinate frame used to formulate the Bubble Ball problem.}
  \label{fig:bubble_description}
\end{figure}
Considering the coordinate frame as shown in Fig.~\ref{fig:bubble_description}, we have:
\begin{align}
    s_k = [x_{\mathrm{ball}}, y_{\mathrm{ball}}, \dot{x}_{\mathrm{ball}},  \dot{y}_{\mathrm{ball}}]^\top \in \mathbb{R}^4,~ m_k=[(s_{\mathrm{obj},k}^{1})^\top,\dots,(s_{\mathrm{obj},k}^{n})^\top]^\top \in \mathbb{R}^{8\times n},
\end{align}
where the state of the given $i^{\mathrm{th}}$ movable object is denoted as $s_{\mathrm{obj}}^i$.
The state of each object contains its position, angle with the positive x-axis, velocity, angular velocity, width, and height\footnote{The mass and the moment of inertia are calculated from the width and the height, since a constant, known object density is used in this game.}, i.e.,
\begin{align}
    & s_{\mathrm{obj}}^{i} = [x_{\mathrm{obj}}^{i}, y_{\mathrm{obj}}^{i}, \theta_{\mathrm{obj}}^{i}, \dot{x}_{\mathrm{obj}}^{i}, \dot{y}_{\mathrm{obj}}^{i}, \dot{\theta}_{\mathrm{obj}}^{i}, w_{\mathrm{obj}}^{i}, h_{\mathrm{obj}}^{i}]^\top,~\forall i \in \{1,2,\dots,n\}.
\end{align} 
The immovable objects' vector $z$ consists of the ground blocks' states $s_\mathrm{grd}$, with the same components as $s_{\mathrm{obj}}$, i.e,  $z = [(s_{\mathrm{grd}}^1)^\top, \dots , (s_{\mathrm{grd}}^p)^\top]^\top \in \mathbb{R}^{8 \times p}$.
In equation \eqref{eq:basic_sys}, $f(\cdot, \cdot)$ is obtained by discretizing the Euler-Lagrangian equation, which includes contact forces between the ball and the objects.
We assume that $s_0$ is fixed. Our goal is to decide an initial state $m_0$, such that $s_k$ reaches a target set $\mathcal{T}_{\mathrm{tar}}$ at any time step $\tau \leq T$ during evolution, i.e., $\exists \tau \in \{0,1,\dots, T\}: s_\tau \in \mathcal{T}_{\mathrm{tar}}$, where $T>0$ is the maximum simulation time.
Thus, our goal is formulated as a feasibility problem as:  
\begin{subequations}\label{eq:goal_feasilbity}
\begin{align}
    & ~~~\min_{m_0}~~ 1,  \nonumber \\
    &~~~~~\mathrm{s.t.,} ~ \begin{bmatrix} s_{k+1} \\ m_{k+1} \end{bmatrix} = \begin{bmatrix}f([s_k^\top,m_k^\top]^\top,z)\\g([s_k^\top,m_k^\top]^\top,z)\end{bmatrix},  \label{dyn_opt2} \\
    & ~~~~~~~~~~~~ s_0 = \bar{s}_0,  \nonumber \\
    & ~~~~~~~~~~~~ s_k \in \mathcal{S}(z) , m_k  \in \mathcal{M}(z),~\forall k \in \{0,1,\dots,T\}, \label{con1_opt2}\\
    & ~~~~~~~~~~~~ \exists \tau \in \{0,1,\dots, T\}: s_\tau \in \mathcal{T}_\mathrm{tar}, \label{con2_opt2}
\end{align}
\end{subequations}
where $\mathcal{S}(z), \mathcal{M}(z)$ are the set of constraints on the ball's and the movable objects' states, respectively, imposed by the environment, and $\bar{s}_0$ is a known constant vector. Note that we only consider taking the position of the ball to a small neighborhood of a target position $(x_\mathrm{tar}, y_\mathrm{tar})$ without any conditions on the target velocity. Therefore the set $\mathcal{T}_\mathrm{tar}$ can be defined as:
\begin{align}\label{eq:target_set_1}
       \mathcal{T}_{\mathrm{tar}} = \{s:s \in \mathbb{R}^4,  \Big|\Big|\begin{bmatrix}s^{(1)} \\ s^{(2)}\end{bmatrix} - \begin{bmatrix}x_\mathrm{tar} \\ y_\mathrm{tar}\end{bmatrix} \Big|\Big|_2 \leq \epsilon \},~\textnormal{for some $\epsilon >0$},
\end{align}
where $v^{(i)}$ denotes the $i^\mathrm{th}$ component of any vector $v$.
\begin{assumption}[Well Posedness]\label{assump:well_posed}
We assume that there exists a feasible solution to problem \eqref{eq:goal_feasilbity}.
\end{assumption}
There are two main challenges in solving \eqref{eq:goal_feasilbity}, namely: $(i)$ time horizon $T$ can be large, making computations cumbersome, and $(ii)$ the dynamics in \eqref{dyn_opt2} are affected by contact forces and friction, whose magnitude, direction and spatio-temporal characteristics are hard to predict. One way of resolving $(ii)$ is to build high-fidelity models of all the dynamics equations in  
\eqref{eq:goal_feasilbity},
and then learn its parameters (basically reverse engineering the simulator). This would be very hard and would require  solving a Linear Complimentarity Problem at every time step~\cite{todorov2012mujoco,hwangbo2018per}. To address these challenges, we propose a hierarchical framework to obtain a feasible solution to \eqref{eq:goal_feasilbity}, focusing on a small problem locally and iteratively updating the local problem using observed trajectories of the ball. 

In the next section, we outline the key principles that we use in our proposed approach to find a solution to \eqref{eq:goal_feasilbity}. These principles are then elaborated further in Section~\ref{sec:proposed_approach_detail}, where we finally pose our tractable approximation to \eqref{eq:goal_feasilbity} in local partitions of the state-space. 


\section{Proposed Approach Outline}
In order to simplify the formalism of the proposed approach (which can be very notation-heavy when properly formalized) we next outline the three  main principles used in our solution.\\
\textbf{(i) Locality.} \\
At each trial the solver restricts its focus to a region of the state-space called the \emph{Local Region}, where a candidate solution to \eqref{eq:goal_feasilbity} requires improvement (as it has failed at the previous iteration). By doing so, we reduce the computational complexity associated with \eqref{eq:goal_feasilbity}. 
\\
\textbf{(ii) Iterative Trials.} \\
A simulation of the game according to dynamics \eqref{eq:basic_sys} is referred to as an iteration or trial. The data from an iteration is subsequently utilized to improve the candidate solution in the Local Region and decide a next Local Region. 
\\
\textbf{(iii) Hierarchical Decomposition.} \\
Three hierarchies are used in our solution and can be summarized as: \\
\textbf{H-1: High-Level Guide Path}: In this hierarchy we plan a path of the ball from the starting position of $\bar{s}_0$ to the target set $\mathcal{T}_\mathrm{tar}$ using an off-the-shelf planner, such as Probabilistic Road Map, RRT, RRT*. This path is called the Guide Path, and it only uses the geometry of the free space. Based on the deviation of a recorded ball trajectory from the Guide Path, we identify a Local Region, where objects need to be placed. \\
\textbf{H-2: Kinematics and the Event Dependent Model}: Using simple kinematics of the objects and the environment, we derive a set of prediction models describing the ball's state evolution. Each one of such  models (referred to as  $\bar{f}^\mathrm{kin}(\cdot, \cdot)$ in Section~\ref{subsec:hier2}) is a function of the objects' state or the environment state which are predicted to enter in contact with the ball.

\textit{Main and Supporting Blocks}.
Inside the Local Region, the solver focuses on placing only \emph{one} object which will enter in contact with the ball. We refer to such an object as the \emph{Main Block}. The Main Block states are to be chosen to lower the deviation of the ball's predicted path from the Guide Path. The objects that constrain the position and velocity of the Main Block to the desired ones, are called the \emph{Supporting Blocks}. 

\textit{Event Dependent Optimization}. The  prediction model of the ball changes depending on the contact status, i.e., if the ball has  entered in contact with some objects or the environment. 
For instance, the ball can enter in contact with an object and then roll on it or can bounce off to another object.
We call each instance of such change an \emph{event}. We model the ball's transition between any two such events in a Local Region, using model $\bar{f}^\mathrm{kin}(\cdot, \cdot)$. This enables a fast prediction as a function of the Main Block's state. \\
\textbf{H-3: Model Learning from Iteration Data}: After the Main and Supporting Block states are chosen in a Local Region, we simulate the game according to \eqref{eq:basic_sys} and record the ball's trajectory. This trajectory data is used to update the prediction models $\bar{f}^\mathrm{kin}(\cdot, \cdot)$ and refine the solution. The next Local Region is chosen as per H-1 using this trajectory and H-2 is repeated.

\section{Proposed Approach Details}\label{sec:proposed_approach_detail}
In this section we outline the optimization problem solved at each hierarchical level discussed in the previous section.

\subsection{Hierarchy 1: Guide Path and Local Region}\label{subsec:hier1}
\begin{figure}[H]
\centering
\subfloat[Guide Path $\Gamma_g$ (blue circles), recorded trajectory $\Gamma_\mathrm{in}$ (red circles), Local Region $\mathcal{R}^\mathrm{loc}(\Gamma_\mathrm{in}, \Gamma_g)$ (red rectangle), and the Local Target set $\mathcal{T}^\mathrm{loc}_\mathrm{tar}$ (blue disc) ]{\includegraphics[width=0.43\columnwidth]{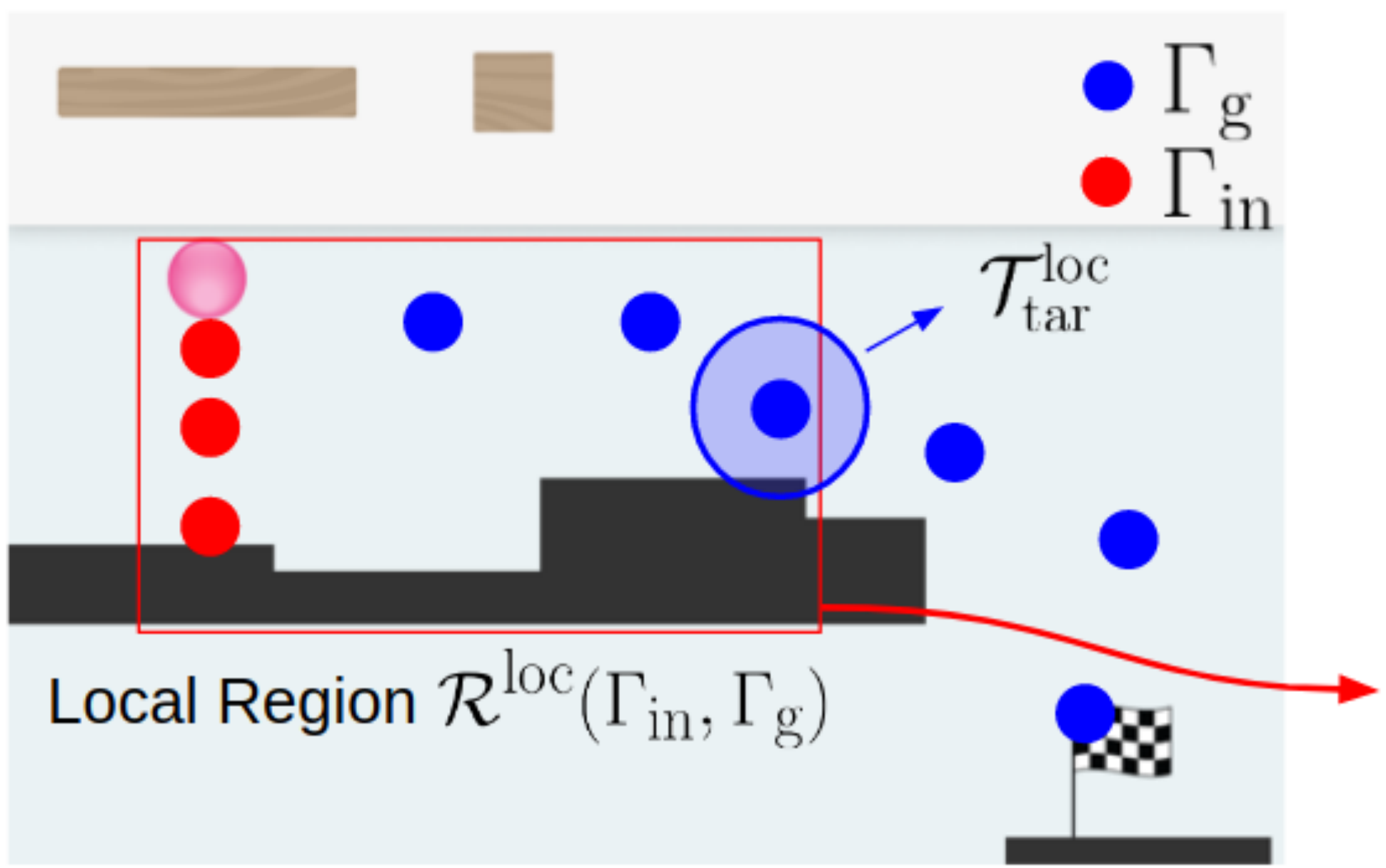}\label{fig:localregion}}
\hspace{0.5cm}
\subfloat[Main Block and Supporting Blocks inside the Local Region $\mathcal{R}^\mathrm{loc}(\Gamma_\mathrm{in}, \Gamma_g)$]{\includegraphics[width=0.43\textwidth]{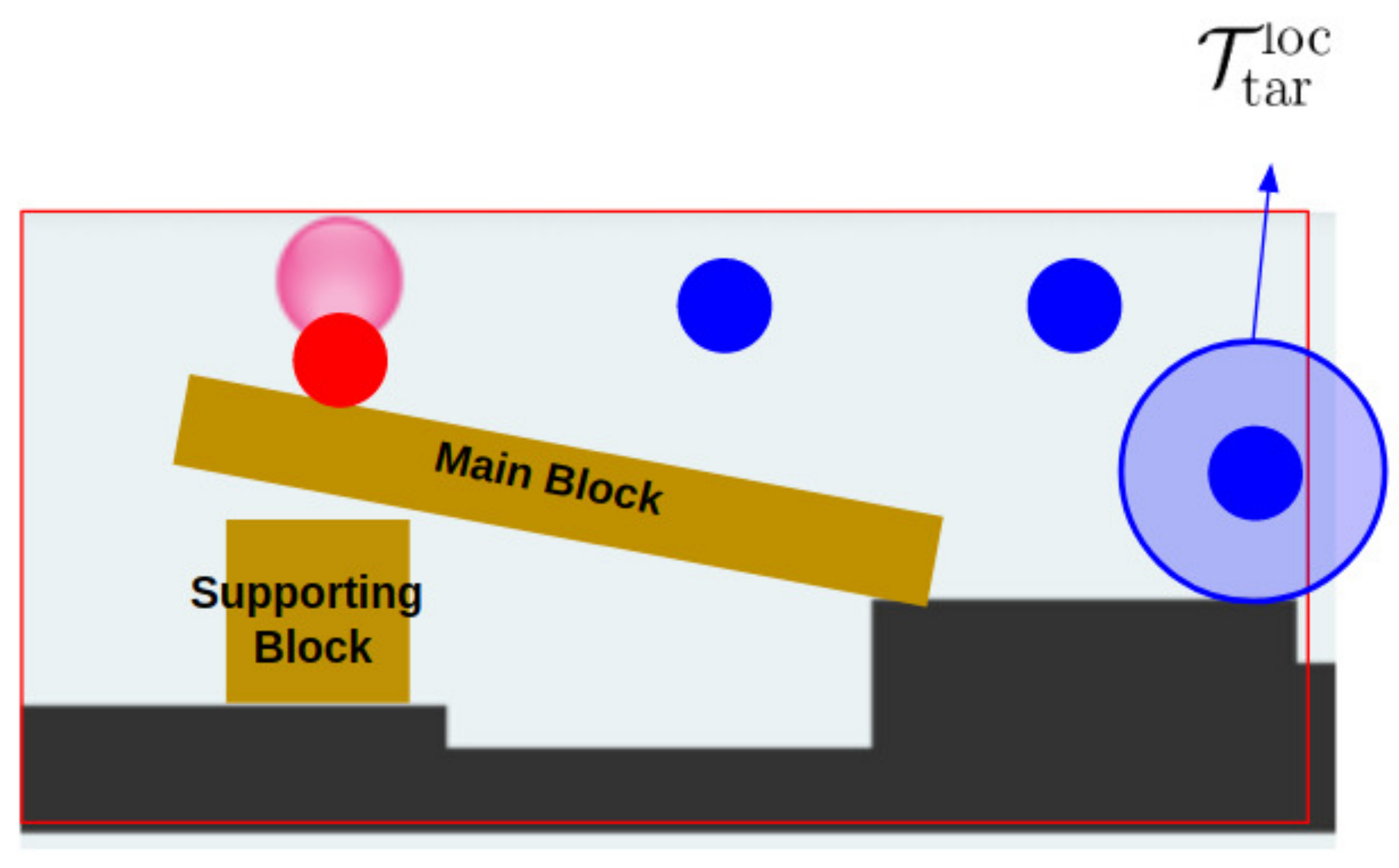}\label{fig:block_and_event}}
\caption{}
\end{figure}

Next we formalize the algorithm at Hierarchy 1 (H-1). Denote by $\Gamma_{\mathrm{in}}$ a recorded trajectory of the ball $\Gamma_{\mathrm{in}} = \{[x_{\mathrm{ball},0},y_{\mathrm{ball},0}]^\top,[x_{\mathrm{ball},1},y_{\mathrm{ball},1}]^\top,\dots,[x_{\mathrm{ball},\tau},y_{\mathrm{ball},\tau}]^\top\}$ at some trial $j$ of the game evolving according to \eqref{eq:basic_sys}, with a given configuration of movable objects (the selection of the movable objects will be discussed in H-2 and H-3 in Section~\ref{subsec:hier2} and Section~\ref{subsec:hier3} later). At the first trial $j=0$ the $\Gamma_{\mathrm{in}}$
is obtained by recording the ball motion with no moving objects placed in the environment. This is shown in Fig.~\ref{fig:localregion} with the red circles. 

Assume $\Gamma_{\mathrm{in}}$ does not solve the game~\eqref{eq:goal_feasilbity}. The goal is to solve the game \eqref{eq:goal_feasilbity} at the next trial $j\!+\!1$.
An approximation of the desired trajectory of the ball to reach the target flag (see Fig.~\ref{fig:bubble_levels}) is denoted by $\Gamma_g$ and is computed by using an off-the-shelf path planner which uses no information about the game kinematics and dynamics. The deviation between $\Gamma_g$ and $\Gamma_{\mathrm{in}}$ is used to identify a region  $\mathcal{R}^\mathrm{loc}(\Gamma_\mathrm{in}, \Gamma_g)$ in the  $(x,y)$ space  where movable objects are to be placed/replaced or moved.
The planned path $\Gamma_g$ is called \textit{Guide Path}, as shown with the blue circles in Fig.~\ref{fig:localregion}, and the region $\mathcal{R}^\mathrm{loc}(\Gamma_\mathrm{in}, \Gamma_g)$ is called a \textit{Local Region}, as shown with the red rectangle in Fig.~\ref{fig:localregion}. 

Let the Guide Path be represented as an ordered set of vectors:
\begin{align*}
    \Gamma_g = \{[x_0^g,y^g_0]^\top, \dots, [x^g_\tau,y^g_\tau]^\top\},
\end{align*}
where the planner trajectory states $\{[x_0^g,y^g_0]^\top, \dots, [x^g_\tau,y^g_\tau]^\top\}$ satisfy the static environment constraints \eqref{con1_opt2}-\eqref{con2_opt2}, but \emph{not} the dynamics \eqref{dyn_opt2}. That is, the Guide Path only uses the geometry of the free space. 
Next we formally define the Local Region $\mathcal{R}^\mathrm{loc}(\Gamma_\mathrm{in}, \Gamma_g)$. 

\begin{definition}[Local Region]\label{def:local_region}
For any $\epsilon_2>\epsilon_1>0$ , a Local region $\mathcal{R}^\mathrm{loc}(\Gamma_\mathrm{in}, \Gamma_g)$
is the smallest rectangle containing all $v_k,u_k$ pair with $v_k\in \Gamma_\mathrm{in}$ and  $u_k\in \Gamma_g$ where $\Vert(v_k-u_k)\Vert_2\in [\epsilon_1,\epsilon_2]$. 
\end{definition}
We define 
$\Gamma_\mathrm{in}^\mathrm{loc}=\Gamma_\mathrm{in}\cap\mathcal{R}^\mathrm{loc}(\Gamma_\mathrm{in}, \Gamma_g)$
and $\Gamma_{g}^\mathrm{loc}=\Gamma_{g}\cap\mathcal{R}^\mathrm{loc}(\Gamma_\mathrm{in}, \Gamma_g)$.


At this hierarchical level, the problem of solving \eqref{eq:goal_feasilbity} is converted into  solving a simpler optimization problem (over the initial state of the movable block) in the Local Region, trying to place the ball in a Local Target Set.
A Local Target set is a neighbourhood of the last element of the guide path 
$\Gamma_g^{\mathrm{loc}}(\mathrm{end})$ in the Local Region, as shown in with the blue disc in Fig.~\ref{fig:localregion}. Here, for a  vector (ordered set of vectors)  $v$, $v\mathrm{(end)}$ refers to the   last component (vector) of $v$.

\begin{definition}[Local Target Set]\label{def:local_target_set}
For a given $\epsilon >0$ we define a Local Target Set as:
\begin{align}\label{eq:local_target_set}
    & \mathcal{T}_{\mathrm{tar}}^{\mathrm{loc}} = \{s \in \mathbb{R}^4: ||\Gamma_g^{\mathrm{loc}}(\mathrm{end})-[s^{(1)}, s^{(2)}]^\top||_2 < \epsilon\}.
\end{align}
\end{definition}

In summary, at this hierarchical level (H-1): $(i)$ a $\Gamma_g$, called the Guide Path is generated, $(ii)$
given $\Gamma_\mathrm{in}$, i.e., the recorded ball trajectory from last trial, and $\Gamma_g$, the Guide Path,
we compute a Local Region and a Local Target Set, and  
$(iii)$ problem \eqref{eq:goal_feasilbity} is then transformed into the following problem which  attempts to place the ball in the Local Target Set
while following the Guide Path: 
\begin{equation}\label{eq:local_opt}
\begin{aligned}
    &\min_{k^\mathrm{loc}, m_{k^\mathrm{loc}}} ~~~~\ell(s_{T^{\mathrm{loc}}}, \Gamma_g^\mathrm{loc}(\mathrm{end}))  \\
    &~~~~~~\mathrm{s.t.,} ~~~~~ s_{k+1}  = f(\begin{bmatrix} s_k \\ m_k \end{bmatrix}, z),  \\
    & ~~~~~~~~~~~~~~~~m_{k+1}  = g(\begin{bmatrix} s_k \\ m_k \end{bmatrix} , z),  \\
    & ~~~~~~~~~~~~~~~~ s_{k^\mathrm{loc}} = \Gamma_{\mathrm{in}}^\mathrm{loc}(k^\mathrm{loc}),  \\
    & ~~~~~~~~~~~~~~~~  s_k \in \mathcal{S}(z) ,
    ~m_k  \in \mathcal{M}(z),~\forall k \in \{k^\mathrm{loc},k^\mathrm{loc}\!+\!1,\dots,T^{\mathrm{loc}}\},  \\
\end{aligned}
\end{equation}
where $\ell(\cdot, \cdot)$ is a positive definite cost function,  $T^{\mathrm{loc}} \ll T$ and $k=k^{\mathrm{loc}}$ is the time instant when the  ball's trajectory $\Gamma_\mathrm{in}$ requires a change of course through impact with an object. Here the optimization is formulated over $k^{\mathrm{loc}}$ and the initial states of the selected objects at time step $k^{\mathrm{loc}}$. Spatio-temporal locality helps focusing on the small problem \eqref{eq:local_opt} instead of \eqref{eq:goal_feasilbity}. Since the original problem \eqref{eq:goal_feasilbity} is to find the objects' state at time step $k=0$, we reverse calculate $m_0$ from $m_{k^\mathrm{loc}}$ in \eqref{eq:local_opt}, and assume this is always possible. 

Even though we reduced computational complexity by focusing inside the Local Region, still the solution to \eqref{eq:local_opt} is complex due to the unknown/complex  $f(\cdot, \cdot)$ and $g(\cdot, \cdot)$. 
Therefore, problem \eqref{eq:local_opt} is not solved at this hierarchy.
The proposed algorithm instead builds a simple approximation to model $f(\cdot,  \cdot)$ in the next hierarchy. This is done by focusing on only one object affecting the ball and using kinematic modeling. At this hierarchy H-2 the algorithm finds a feasible solution to \eqref{eq:local_opt} using this simple model approximation.  





\subsection{Hierarchy 2: Kinematics and the Event Dependent Model
}\label{subsec:hier2}

In the Local Region, we 
allow only one object at a time (including the immovable objects) to affect the ball's motion. We call the movable object which will affect the ball's motion in the Local Regions as the \textit{Main Block}. This is shown in Fig.~\ref{fig:block_and_event}, where the rectangular movable block is the Main Block inside the chosen Local Region.

Consider the Main Block as the $l^\mathrm{th}$ object out of the $n$ objects, and denote its state at time $k$ as $ m^{l}_k$, which is a vector stacking $6(l-1)^{\mathrm{th}}$ to the $6l^{\mathrm{th}}$ element of $m_k$. Additionally, the objects used to constrain the states of the Main Block to the desired one are called the \textit{Supporting Block}s. In Fig.~\ref{fig:block_and_event} the square movable block is the Supporting Block, which ensures that the Main Block stays pivoted from the left, and maintains its chosen angle. 
Then, we can rewrite the dynamics from \eqref{eq:basic_sys}:
\begin{align}\label{eq:pick_one_f}
    s_{k+1} = \bar{f}(s_k, m_k^{l}),
\end{align}
and 
since we focus on only $m^{l}_k$, we can also take only $m^{l}_k$ from $g(\cdot, \cdot)$. We can write this as:
\begin{align}\label{eq:pick_one_g}
    m^{l}_{k+1} = \bar{g}(\begin{bmatrix} s_k \\ m_k \end{bmatrix} , z)
\end{align}

\begin{remark}
Argument $m^{l}_k$ in \eqref{eq:pick_one_f} is substituted by $z^{q}$ when the $q^\mathrm{th}$ immovable object affects the ball. 
\end{remark}

\begin{remark}
When no object contacts the ball, $f(\cdot, \cdot)$ can be $\bar{f}_g(s_k)$, evolving as a free-fall. 
\end{remark}


\subsubsection*{Kinematics Based Approximation}
Although by using \eqref{eq:pick_one_f} and \eqref{eq:pick_one_g}, we reduce the complexity of the system evolution function in \eqref{eq:local_opt}, we still need to solve the contact dynamics for the ball and the object in order to predict the ball's motion. Instead, we use kinematics based approximation of $\bar{f}(\cdot,\cdot),~\bar{g}(\cdot,\cdot)$. This allows us to predict the ball's trajectory across the Local Region to solve a tractable approximation to problem \eqref{eq:local_opt}. 

In order to do so, we classify $C$ different possible types of contacts between the ball and the object, i.e., contacts involving ball and a line, ball and a circle, and so on. For each such type of interaction, we construct a kinematic model for the corresponding trajectories such as the ball moving along the line, changing the direction after a collision, etc. This is similar to the notion of building motion primitives, except that we consider contacts.

We notice that while the $l^\mathrm{th}$ object, i.e, the Main Block, is in contact with the ball, the variation of it's state  $m^{l}$ is negligible in the Bubble Ball game. With this observation, model \eqref{eq:pick_one_g} is not needed to compute the evolution of \eqref{eq:pick_one_f} while the ball is in contact with the Main Block. 
In summary, we can write a kinematic approximation to \eqref{eq:pick_one_f} for a type $j$ interaction with the Main Block $l$ as:
\begin{align}\label{eq:approx_kin}
    s_{k+1} =  \bar{f}^{\mathrm{kin},j}(s_k,m^{l}_k),~\textnormal{for some } j \in \{1,2,\dots,, C\}.
\end{align}
By using this approach we obtain a tractable approximation to the optimization problem \eqref{eq:local_opt} as:
\begin{equation}\label{eq:local_kin_opt}
\begin{aligned}
    &\min_{k^\mathrm{loc},m_{k^\mathrm{loc}|k^\mathrm{loc}}} ~~~~\ell(s_{T^{\mathrm{loc}}|k^\mathrm{loc}}, \Gamma_g^\mathrm{loc}(\mathrm{end}))  \\
    &~~~~~~~\mathrm{s.t.,} ~~~~~~~~
    s_{k+1|k^\mathrm{loc}}  = \begin{cases}\bar{f}^{\mathrm{kin},j_l}(s_{k|k^\mathrm{loc}},  m^{l}_{k|k^\mathrm{loc}}), & \textnormal{while}~ m^{l}_{k|k^\mathrm{loc}} ~\textnormal{is active} \\
    \bar{f}^{\mathrm{kin},j_q}(s_{k|k^\mathrm{loc}},  z^{q}), & \textnormal{while}~ z^{q} ~\textnormal{is active}
    \\
    \bar{f}_g(s_{k|k^\mathrm{loc}}), & \textnormal{otherwise}~~ \end{cases},  \\
    & ~~~~~~~~~~~~~~~~~~~~ s_{k^\mathrm{loc}|k^\mathrm{loc}} = \Gamma_{\mathrm{in}}^\mathrm{loc}(k^\mathrm{loc}),~m^l_{k+1|k^\mathrm{loc}}=m^l_{k|k^\mathrm{loc}},~\textnormal{while } m^{l}_{k|k^\mathrm{loc}} ~\textnormal{is active},  \\
    & ~~~~~~~~~~~~~~~~~~~~  s_{k|k^\mathrm{loc}} \in \mathcal{S}(z), m_{k|k^\mathrm{loc}}  \in \mathcal{M}(z),~\forall k \in \{k^\mathrm{loc},k^\mathrm{loc}\!+\!1,\dots,T^{\mathrm{loc}}\},\\
    & ~~~~~~~~~~~~~~~~~~~~ l \in \{1,2,\dots,n\}, 
\end{aligned}
\end{equation}
where $s_{k|k^{\mathrm{loc}}}$ is the predicted ball state at time step $k$ according to the model $\bar{f}^{\mathrm{kin},j}(\cdot,\cdot)$, and $m_{k|k^{\mathrm{loc}}}^{l}$ is the predicted Main Block's state at time step $k$, when the predictions are made given the ball's recorded position at time step $k^{\mathrm{loc}}$,  $\Gamma^\mathrm{loc}_\mathrm{in}(k^\mathrm{loc})$ . 


We now transform problem  \eqref{eq:local_kin_opt} 
into an \emph{Event-Dependent} optimization by exploiting the simple nature of the contacts in the game. This lowers the computational complexity of solving \eqref{eq:local_kin_opt} and also allows us to pinpoint where the switches in the ball's prediction model occur.

\subsubsection*{Event Based Simplification}
An \emph{event} is the moment when the contact constraint on the ball changes. That is, an event occurs when the ball's motion model in \eqref{eq:local_kin_opt} switches. For instance, in case of a free-falling ball hitting a block and rolling thereafter, an event is the collision with the block.

We define the \emph{Event-Dependent} system as the state transition function 
which  links the ball state from one event to  the next event. More formally,
let $e_{i}$ be the time step when an event $i$ involving $m_{e_i}^{l}$ happens. Then the corresponding Event-Dependent model $\bar{f}^{\mathrm{kin},j}_{e_i}$ is denoted by:
\begin{align}\label{eq:event_dep_kin}
    s_{e_{i+1}} = \bar{f}^{\mathrm{kin},j}_{e_i}(s_{e_i},m_{e_i}^{l}),~\textnormal{for some } j \in \{1,2,\dots,, C\}.
\end{align}
\begin{remark} \label{rem:simple_int}
To obtain $\bar{f}^{\mathrm{kin},j}_{e_i}(\cdot,\cdot)$ in \eqref{eq:event_dep_kin} from $\bar{f}^{\mathrm{kin},j}(\cdot,\cdot)$ in \eqref{eq:local_kin_opt}, we simply forward integrate f $\bar{f}^{\mathrm{kin},j}(\cdot,\cdot)$ over time in advance. In case of rolling, the integration stops when the motion reaches the object $m^l$ boundary. In case of bouncing at time step $e_i$, the instantaneous value of $\bar{f}^{\mathrm{kin},j}(\cdot,\cdot)$ at $k=e_i$ is used as $\bar{f}^{\mathrm{kin},j}_{e_i}(\cdot,\cdot)$. The bouncing then switches to free-fall at $k=e_i+1$. The same principle of forward  integration as in rolling is also used for a free-falling motion using $\bar{f}_g(\cdot)$ and it is denoted as $\bar{f}_{g,e_i}(s_{e_i},d)$ with x-distance $d$ to the next event. 
\end{remark}
Using \eqref{eq:event_dep_kin} we transform the time-dependent problem \eqref{eq:local_kin_opt} into an event-based optimization over a single event for one Main Block in each Local Region. In each Local Region, we limit the search over  at most three events, namely $(i)$ event 1: the ball enters in contact with the Main Block at time step $e_1$, $(ii)$ event 2: the ball leaves the Main Block and goes in free-falling at time step $e_2$ $(iii)$ event 3: the ball enters in contact with the environment at time step $e_3$, until $e_4$. Now, for event 1, Main Block is the only one affecting the ball, and after event 1 ends, only the environment  $z^{q}$ plays a role in the ball's evolution. 
With these notations, the reformulated optimization problem becomes:
\begin{equation}\label{eq:opti_bubble}
    \begin{array}{llclcl}
\displaystyle \min_{e_1,l,m^{l}_{e_1}} & \multicolumn{3}{l}{\ell(s_{{e}_4|e_1}, \Gamma_g^\mathrm{loc}(\mathrm{end}))} \\
~~~\textrm{s.t.,} &  s_{{e}_1|e_1} & = & \Gamma_{\mathrm{in}}^{\mathrm{loc}}(e_1),~\textnormal{(note that $e_1 = k^\mathrm{loc}$)} \\
&\displaystyle s_{{e}_2|e_1} & = & \bar{f}^{\mathrm{kin},j_l}_{e_1}( s_{{e}_1|e_1},m^{l}_{e_1}),~\textnormal{($j_l$ is known for $l$)} \\
& s_{{e}_3|e_1} & = & \bar{f}_{g,e_2}(s_{{e}_2|e_1},d),   \\
&\displaystyle s_{{e}_4|e_1} & = & \bar{f}^{\mathrm{kin},j_q}_{e_3}( s_{{e}_3|e_1},z^{q}),~\textnormal{($j_q$ is known for $q$)}\\
&\displaystyle s_{{e}_i|e_1} & \in & \mathcal{S}(z),~
\forall i \in \{1,2,3,4\}\\
&\displaystyle m_{{e}_1|e_1} & \in & \mathcal{M}(z),

\end{array}
\end{equation}
where $s_{e_i|e_1}$ is the predicted ball state at event $i$ using $\bar{f}_{e_i}^{\mathrm{kin},j}(\cdot,\cdot)$, given $e_1$, and $m_{e_i|e_1}$ is the predicted Main Block's state at event $i$. 
This means solving \eqref{eq:opti_bubble} in the Bubble Ball is equivalent to $(i)$ finding the best {Main Blocks} $l$, and $(ii)$ finding $ m^{l}_{e_1}$ with the corresponding {Supporting Blocks} states. 
where $z^{q}$ is the given ground's state and $d$ is the distance to the given ground. 
The decision variables are the starting time of event 1 (i.e., time of contact with the Main Block) ($e_1$), choice of the Main Block ($l$), the Main Block's state at contact ($m^{l}_{e_1}$).

\begin{remark}\label{rem:grid}

To solve the optimization problem \eqref{eq:opti_bubble}, we use a random shooting method for several potential Main Blocks indices $l$.
We first choose a Coarse Gridding (CG) of the decision space and choose the best block $l^\star$ that minimizes the cost in \eqref{eq:opti_bubble}. With this $l^\star$ (i.e., the chosen Main Block), we then repeat a Finer Gridding (FG) of the remaining decision variables. The combination yielding the minimum cost is then chosen as an approximate solution to \eqref{eq:opti_bubble}.
\end{remark}

\subsubsection*{From Kinematic to Dynamic Modeling}
In our algorithm,  we also need dynamic modeling including  moment equilibrium and rotational dynamics to predict $\bar{g}(\cdot, \cdot)$ in \eqref{eq:pick_one_g}. We use such information in order to choose the Supporting Blocks' location, corresponding to $m_{e_1}^l$ from the solution in \eqref{eq:opti_bubble}. 
Since the contact forces to the Main Block are applied on multiple points, we cannot easily approximate the evolution of the Main Block's state using only kinematics. Moreover, if $m^l_{e_1}$ has a non-zero velocity (e.g., \href{https://www.youtube.com/watch?v=bA0LEdjjWeY&feature=emb_logo}{catapult}), we need to predict the state of the Main Block with time-varying contact forces and it requires dynamics. We also need some iterations until we find correct initial state corresponding to $m^l_{e_1}$ since we do not know the perfect $\bar{g}$.  We do not present this technical detail in this paper. Note that currently this is the only feature where we are using dynamics information of the objects, along with kinematics. After solving \eqref{eq:opti_bubble}, we pick the Main Block's and corresponding Supporting Blocks' states in the Local Region and simulate the game according to \eqref{eq:basic_sys}. 

Due to the inherent simplifications made in the models \eqref{eq:event_dep_kin}, the solution to \eqref{eq:opti_bubble} (when reported to the initial time step 0) may not solve the original problem \eqref{eq:goal_feasilbity}, thus resulting in a failed trial. To alleviate this, in H-3 we use data-driven learning to update the models \eqref{eq:event_dep_kin} and refine our solution iteratively using data from these trials. 
 
\subsection{Hierarchy 3: Model Learning From Iteration Data
}\label{subsec:hier3}
In Hierarchy 3, we focus on data-driven iterative improvement of the kinematic models \eqref{eq:event_dep_kin} derived in Hierarchy 2. Once \eqref{eq:opti_bubble} is solved and the Supporting Blocks are chosen, we simulate the game engine according to \eqref{eq:basic_sys}. We obtain a set of transition data from this trial, denoted as  $[s_{e_i},m^{l}_{e_i}]^{i=N_e}_{i=1}$ from $N_e$ Events along the entire collected trajectory. The models $\bar{f}^{\mathrm{kin},j}_{e_i}(\cdot, \cdot)$ in \eqref{eq:opti_bubble} are parametrized with a vector of parameters $\beta$ for each model. Then, we update $\beta$ as: 
\begin{align}\label{eq:regression}
    \beta \leftarrow \arg \min_\beta \sum_{i=1}^{N_e} |Y_i-\bar{f}^{\mathrm{kin},j}_{e_i}(s_{e_i},m_{e_i}^{l},\beta)|^2,
\end{align}
where $Y_i$ is data of $s_{e_{i+1}}$ from observations along the collected ball trajectory. 


\begin{remark}\label{rmk:iteration}
 The process explained in Hierarchy 1 is repeated after we follow the algorithms outlined in Hierarchy 2 and 3 and find a solution trajectory
 $\Gamma^\star$  in the current Local Region. Recall $\Gamma_\mathrm{in}$ from Section~\ref{subsec:hier1}. We reset $\Gamma_{\mathrm{in}} \leftarrow \Gamma^\star$ and repeat the process until the game is solved. 
\end{remark}

\section{Analysis of the Results}\label{sec:result}
We used our proposed approach to successfully solve a set of Bubble Ball levels. For many of the levels we solved, the simple kinematics based models from Hierarchy 2 captured the featured motions, and only a few of iterations in Hierarchy 3 were enough to refine the solution.
Our \href{https://sites.google.com/berkeley.edu/bubble-ball/home}{BubbleBall Website}
gives a visual explanation of this proposed hierarchical approach and shows how it 
works under one or multiple Local Regions, and even a dynamic Main Block (e.g, a catapult in level 21). 
In Table~\ref{table:results}, we show the number of trials needed to solve a set of levels, along with the average computation times of \eqref{eq:opti_bubble} in each Local Region. 
For the solved levels, up to 5 iterations were sufficient, implying that the hierarchy and the kinematics approximation is able to capture the complex game dynamics.   
Also, we have failures on some levels.
 There are three main reasons of such failures. 
In some cases, the simple forward integration model in Remark \ref{rem:simple_int} does not hold due to the intermediate interruption of another block. Second, sometimes wrong Supporting Blocks are chosen, or the corresponding $m_0$ is not chosen well because it is not perfect to predict $\bar{g}(\cdot,\cdot)$. And third, the Guide Path can be greedy, so error minimization from this guide path in the Local Region does not necessarily guarantee success in the level. Each of these components will be improved in our future work. 
\begin{table}[th!]
\begin{tabular}{ |c||p{2.8cm}|p{2.8cm}|c|p{2.3cm}|p{2.3cm}|  }
 \hline
 Level&  Time to Solve \eqref{eq:opti_bubble} with CG (s) &  Time to Solve \eqref{eq:opti_bubble} with FG (s) & No. of Trials & No. of Local Regions& No. of Movable Blocks\\
 \hline
  1  & 0.071  &0.045 &   1 & 1&1\\
 \hline
  2 & 0.069   &0.049&   1 & 1&2\\
 \hline
 3 &  6.541 & 5.583  &1 & 1 & 2\\
 \hline
 4 &3.345 & 2.425&  1 & 1& 2\\
 \hline
 5    &0.672 & 0.539&  2 & 2& 2\\
 \hline
 6&    3.405 & 0.331 & 1 & 1&2\\
 \hline
  8&    0.667  & 0.471 & 5 &3 & 6\\
 \hline
  9&    0.185  & 0.297 & 2 & 2& 4\\
 \hline
  11&   0.625  & 0.262 & 4 &4 & 4\\
 \hline
  21&    5.661  & 0.071 & 4 & 1&5\\
 \hline
\end{tabular}
\caption{Empirical results of the solved levels}
\label{table:results}
\end{table}

\section{Conclusion}
We proposed a hierarchical framework to solve the physics-based game Bubble Ball. Our approach exploits model hierarchy and model learning from iteration data. Hierarchy 1 uses a  geometry-based high-level planner for planning a Guide Path of the ball. 
In Hierarchy 2, using low-level kinematic and dynamic modeling of the blocks, we solved an \emph{Event-Dependent} optimization problem which decides how the game configuration is chosen for the ball to track the Guide Path.
In Hierarchy 3, using the collected game data 
we improve our solution of Hierarchy 2.
Our method successfully solved a large set of levels within reasonable number of trials.  

\section*{Acknowledgements}
We thank Robert Nay for providing access to the simulator and license of Bubble Ball. This research work is partially funded by grants ONR-N00014-18-1-2833 and NSF-1931853.

\bibliographystyle{unsrt}
\bibliography{mybib.bib}

\end{document}